\documentclass[fleqn,twoside]{article}
\usepackage{latexsym}
\usepackage[dvips]{graphicx}
\usepackage{espcrc2}

\author{Matthew A. Nobes\address[MCSD]{Physics Department, Simon
Fraser University Burnaby, B.C., Canada V5A 1S6},
Howard D. Trottier\addressmark,
G. Peter Lepage\address{Newman Laboratory of Nuclear Studies, Cornell
University, Ithaca, NY, 14853},
Quentin Mason\addressmark}

\title{Second Order Perturbation Theory for Improved Gluon and Staggered Quark Actions}

\begin{document}

\begin{abstract}
We present the results of our perturbative calculations of the static
quark potential, small Wilson loops, the static quark self energy, and
the mean link in Landau gauge.  These calculations are done for the
one loop Symanzik improved gluon action, and the improved staggered
quark action.
\end{abstract}

\maketitle

It has long been known that perturbative calculations are necessary
for precision measurements of many quantities.  Unfortunately
perturbation theory using a lattice cutoff is difficult, due to extra
diagrams and vastly more complicated Feynman rules.  The complexity of
these calculations is an impediment to doing the higher--order
perturbation theory required for many important applications of
improved actions.  The present work attempts to automate as much of
the perturbation theory as possible in order to make these types of
computations more straightforward.

For any given action some of the most basic things we would like to
compute are various operators that can be built out of products of
links.  These include the static quark potential, small Wilson loops,
the static quark self energy, and the mean link in Landau gauge.

Perturbative determinations of these quantities are central to many
other applications.  For example, the static quark potential can be
used to determine a physical, renormalized coupling, and the
perturbative expansions of small Wilson loops can be used to extract
the strong coupling from simulations (see \cite{davies95}).  We have
the techniques in place to calculate all of these, for an arbitrary
gluon action, along with the contributions from quark loops for highly
improved quark actions.  This paper reports preliminary results for
the above quantities, for the one-loop Symanzik improved gluon action,
and the improved staggered quark action.

These calculations are similar to traditional continuum perturbative
QCD.  The major difference is that the ultraviolet regulator is the
lattice cutoff, which leads to the computational difficulties
mentioned above.  A further problem is how to regulate infrared
divergences beyond one--loop, in a gauge covariant manner.  This can
be done using twisted periodic boundary conditions \cite{luscher86}.
Feynman rules and factors from the operators are generated using the
L\"uscher and Weisz vertex generation algorithm \cite{luscher86}.
PYTHON or C++ scripts implement this program for a given set of links.
An advantage to this method is the ease with which new actions and
operators can be treated.  We use FORTRAN programs to do the momentum
sums.  These are generally done on large but finite lattices
(typically $100^{4}$ volume).

The action we consider here is the one--loop Symanzik improved action
for isotropic lattices,
\begin{eqnarray}\label{impact}
S_{G} & = & \beta_{pl} \sum_{x;\mu < \nu} (1-P_{\mu\nu}) 
+ \beta_{rt} \sum_{x;\mu \ne \nu} (1-R_{\mu\nu}) 
\nonumber \\
& & + \beta_{pg} \sum_{x;\mu < \nu < \sigma} (1-C_{\mu\nu\sigma}),
\end{eqnarray}
where
\begin{eqnarray}
\beta_{pl} & = & \frac{10}{g^{2}}, \qquad
\beta_{rt}  = -\frac{\beta_{pl}}{20u_{0}^{2}}(1+0.4805\alpha_{s})
\nonumber \\
\beta_{pg} & = & -\frac{\beta_{pl}}{u_{0}^{2}}0.03325\alpha_{s}.
\end{eqnarray}
We present our results for an expansion in the bare lattice coupling
$\alpha_{\rm{latt}}=g^{2}/(4\pi)$. We
include tadpole counterterms for (\ref{impact}), which are generated
by the first order expansion of the mean-field in the rectangle terms.
In this paper we use the mean-field, $u_{0}$, defined by the average
plaquette \cite{bernard98} which, for the action of (\ref{impact}), is
given to first order by $u_{0} \approx 1 - 0.7671 \alpha_{\rm{latt}}.$
Our second order results also included the counterterm generated by
the 1x1x1 paths in (\ref{impact}).

Next we consider the static quark potential which is the central
quantity that needs to be computed.  We can use it to define a
renormalized coupling $\alpha_{V}$, which can be used as the expansion
parameter for other quantities.

To compute the static quark potential we take the correlator of two
Polyakov lines of length $L_{T}$, separated by a distance R,
$<L(R,L_{T})>$.  The static quark potential is then given by
\begin{equation}\label{stat1}
V(R) + 2E_{0} = \lim_{L_{T}\to\infty} \frac{1}{L_{T}} \ln<L(R,L_{T})>,
\end{equation}
where $E_{0}$ is the self energy of an isolated quark (see below for
our calculation of this quantity).  Expanded in
the bare coupling the static quark potential should have the following
form \cite{lepage93}
\begin{eqnarray}\label{stat2}
V(R) & = & - \frac{4}{3} \frac{\alpha_{\rm{latt}}}{R} \nonumber \\
& & \times \lbrace 1 + \alpha_{\rm{latt}}(2\beta_{0}ln(\pi R) + C(R)),
\end{eqnarray}
where $\beta_{0}=(11 - 2/3 n_{f})/(4\pi)$.

As mentioned above, the static quark potential is used to define a
renormalized coupling.  This is done by demanding that the Fourier
transform of (\ref{stat2}) have the form
\begin{equation}
V(q) = -\frac{4}{3} \frac{4\pi}{q^{2}} \alpha_{V}(q).
\end{equation}
Using this definition, we obtain the expansion for the bare coupling
in terms of the physical one,
\begin{eqnarray} \label{barephys}
\alpha_{\rm{latt}} & = & \alpha_{V} \left\lbrace 1 - \alpha_{V} \left( 
2\beta_{0}\ln\left( \frac{\pi}{q} \right) + \widetilde{C} \right)
\right. \nonumber \\
& + & \mathcal{O}(\alpha_{V}^{2}) \bigg\rbrace.
\end{eqnarray}

As a test of our calculations we reproduced the known result for the
Wilson gluon action, $\widetilde{C}=4.702$.  We have also determined
$\widetilde{C}$ for the Symanzik improved gluon action.  We find
\begin{equation}\label{ctilde}
\widetilde{C}_{I} = 3.23(13).
\end{equation}
We could easily recompute $\widetilde{C}$ for other gluon (and quark)
actions, and with other definitions of $u_{0}$.

In addition to the static quark potential, we have computed a number
of other quantities.  Tables \ref{wloop} and \ref{iloop} give results
for the logarithms of small Wilson loops, whose perturbative expansion
is defined by,
\begin{equation}
-\frac{1}{2(R+T)} \ln W(R,T) = \sum_{n} w_{n}(R,T)
 \alpha_{\rm{latt}}^{n}.
\end{equation}
The results for the Wilson action agree with those of \cite{heller85}.

We have also computed the static quark self energy $E_{0}$ through
second order.  We define the self energy $E_{0}(L)$ on a finite
lattice according to
\begin{equation}
E_{0}(L) = -\frac{1}{L} \ln[P_{t}(L)] = \sum_{n} c_{n}(L)
\alpha_{\rm{latt}}^{n}.
\end{equation}
Here $P_{t}(L)$ is the Polyakov line on a lattice of size $L^{4}$.

\begin{figure}[!t]
\caption{The $\mathcal{O}(\alpha_{\rm{latt}})$ contribution to the static quark
self energy.} \label{c1fig}
\begin{center}
\includegraphics[height=7cm,width=5cm,angle=90]{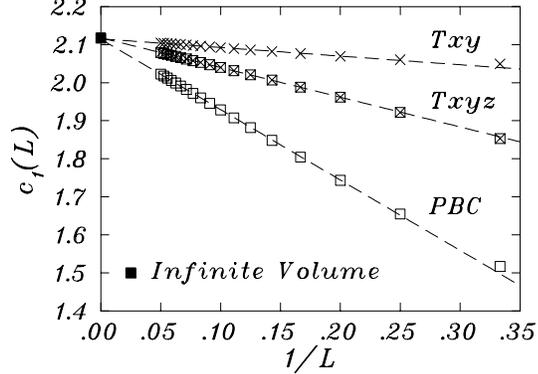}
\end{center}
\end{figure}

\begin{figure}[!t]
\caption{The $\mathcal{O}(\alpha^{2}_{\rm{latt}})$ contribution to the static quark
self energy.} \label{c2fig}
\begin{center}
\includegraphics[height=7cm,width=5cm,angle=90]{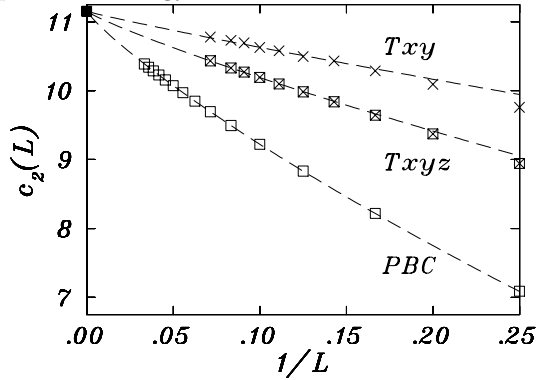}
\end{center}
\end{figure}

Figures \ref{c1fig} and \ref{c2fig} show our results for the first and
second order coefficients on a series of volumes, for three sets of
boundary conditions (Wilson glue, PBC stands for periodic boundary
conditions, Txy and Txyz stand for twisted periodic boundary
conditions along two and three spatial planes, respectively).
The infinite volume extrapolation agree with earlier estimates, $E_{0} = 2.1172\alpha_{\rm{latt}} + 11.152 \alpha^{2}_{\rm{latt}}$.  These
finite volume results were used in a determination of the third order
self energy, using Monte-Carlo methods \cite{lepage2001}.  We have
also calculated the self energy for the improved gluon action (\ref{impact}) with the result: 
$
E_{0} = 1.8347(5) \alpha_{\rm{latt}} + 8.01(2) \alpha_{\rm{latt}}^{2}.
$

Finally, we have computed the mean link in Landau gauge.  In agreement
with earlier determinations we have for the Wilson action,
$
u_{0} = 1 - 0.9738(2) \alpha_{\rm{latt}} - 3.33(1)
\alpha_{\rm{latt}}^{2}.
$
For the one loop Symanzik improved action we report,
$
u_{0} = 1 - 0.7501(1) \alpha_{\rm{latt}} - 2.06(1)
\alpha_{\rm{latt}}^{2}.
$

The methods for automatic vertex generation can also be readily
applied to complicated fermionic actions.  For example, we computed
the $n_{f}$ part of the average plaquette at second order for improved
staggered fermions \cite{lepage99}.  We find:
$
w_{2}(1,1) = 1.958(2) - 0.06969(4) n_{f}.
$
Calculations of the second order $n_{f}$ pieces for the other
quantities considered in this paper are in progress.

We turn now to some conclusions.  Our current results demonstrate the
versatility of our approach.  High precision results, necessary for
accurate determinations of many quantities, are currently being
generated.  For improved staggered quarks
this includes one loop improvement of the action, and the matching of
the quark currents and four quark operators \cite{mackenzie} as well
as quark mass renormalization \cite{hein}.  

\begin{table}
\caption{Perturbative Wilson loops evaluated using Wilson glue, errors
are from the VEGAS integrations.} \label{wloop}
\begin{center}
\begin{tabular}{ c c l l }
\hline
R & T & $w_{1}$ & $w_{2}$ \\
\hline
1 & 1 & 1.0471(4) & 3.548(7) \\
1 & 2 & 1.2041(2) & 4.460(5) \\
1 & 3 & 1.2589(2) & 4.816(6) \\
2 & 2 & 1.4342(3) & 5.841(7) \\
2 & 3 & 1.5177(3) & 6.41(1) \\
3 & 3 & 1.610(1) & 7.09(4) \\
\hline
\end{tabular}
\end{center}
\end{table}

\begin{table}
\caption{Perturbative Wilson loops evaluated using Symanzik improved
glue.} \label{iloop}
\begin{center}
\begin{tabular}{ c c l l }
\hline
R & T & $w_{1}$ & $w_{2}$ \\
\hline
1 & 1 & 0.7673(2) & 1.958(2) \\
1 & 2 & 0.9255(2) & 2.661(3) \\
1 & 3 & 0.9849(2) & 2.954(4) \\
2 & 2 & 1.1503(3) & 3.735(6) \\
2 & 3 & 1.2342(3) & 4.172(8) \\
3 & 3 & 1.3231(4) & 4.666(13) \\
\hline
\end{tabular}
\end{center}
\end{table}

\end{document}